\documentclass[aip,pof,groupedaddress,reprint]{revtex4-1}
\pdfoutput=1
\usepackage{amsmath}
\usepackage{amssymb}
\usepackage{graphicx}
\usepackage{dcolumn}
\usepackage{bm}

\begin{document}

\title[]{Influence of the fluid density on the statistics of power fluctuations in von K\'arm\'an swirling flows}

\author{A. Opazo}
\altaffiliation{Present address: Tecnolog\'ia Integral S.A. Carmen Sylva 2370, Providencia. Santiago, Chile.}
\affiliation{Laboratorio de Turbulencia, Departamento de F\'isica,
Facultad de Ciencia, Universidad de Santiago de Chile, USACH. Casilla
307, Correo 2, Santiago, Chile}

\author{A. S\'aez}
\affiliation{Laboratorio de Turbulencia, Departamento de F\'isica,
Facultad de Ciencia, Universidad de Santiago de Chile, USACH. Casilla
307, Correo 2, Santiago, Chile}

\author{G. Bustamante}
\affiliation{Laboratorio de Turbulencia, Departamento de F\'isica,
Facultad de Ciencia, Universidad de Santiago de Chile, USACH. Casilla
307, Correo 2, Santiago, Chile}

\author{R. Labb\'e}
\email[]{raul.labbe@gmail.com}
\affiliation{Laboratorio de Turbulencia, Departamento de F\'isica,
Facultad de Ciencia, Universidad de Santiago de Chile, USACH. Casilla
307, Correo 2, Santiago, Chile}

\date{16 July 2015}

\begin{abstract}
Here we report experimental results on the fluctuations of injected power in confined turbulence. Specifically, we have studied a von K\'arm\'an swirling flow with constant external torque applied to the stirrers. Two experiments were performed at nearly equal Reynolds numbers, in geometrically similar experimental setups. Air was utilized in one of them and water in the other. With air, it was found that the probability density function of power fluctuations is strongly asymmetric, while with water, it is nearly Gaussian. This suggests that the outcome of a big change of the fluid density in the flow-stirrer interaction is not simply a change in the amplitude of stirrers' response. In the case of water, with a density roughly $830$ times greater than air density, the coupling between the flow and the stirrers is stronger, so that they follow more closely the fluctuations of the average rotation of the nearby flow. When the fluid is air, the coupling is much weaker. The result is not just a smaller response of the stirrers to the torque exerted by the flow; the PDF of the injected power becomes strongly asymmetric and its spectrum acquires a broad region that scales as $f^{-2}$. Thus, the asymmetry of the probability density functions of torque or angular speed could be related to the inability of the stirrers to respond to flow stresses. This happens, for instance, when the torque exerted by the flow is weak, due to small fluid density, or when the stirrers' moment of inertia is large. Moreover, a correlation analysis reveals that the features of the energy transfer dynamics with water are qualitatively and quantitatively different to what is observed with air as working fluid.
\end{abstract}


\maketitle

\section{Introduction}
\label{Intro}

\begin{figure}[t]
\centering \vspace{-0.5cm} \hspace{-0.2 cm}
\includegraphics[width=.75\textwidth]{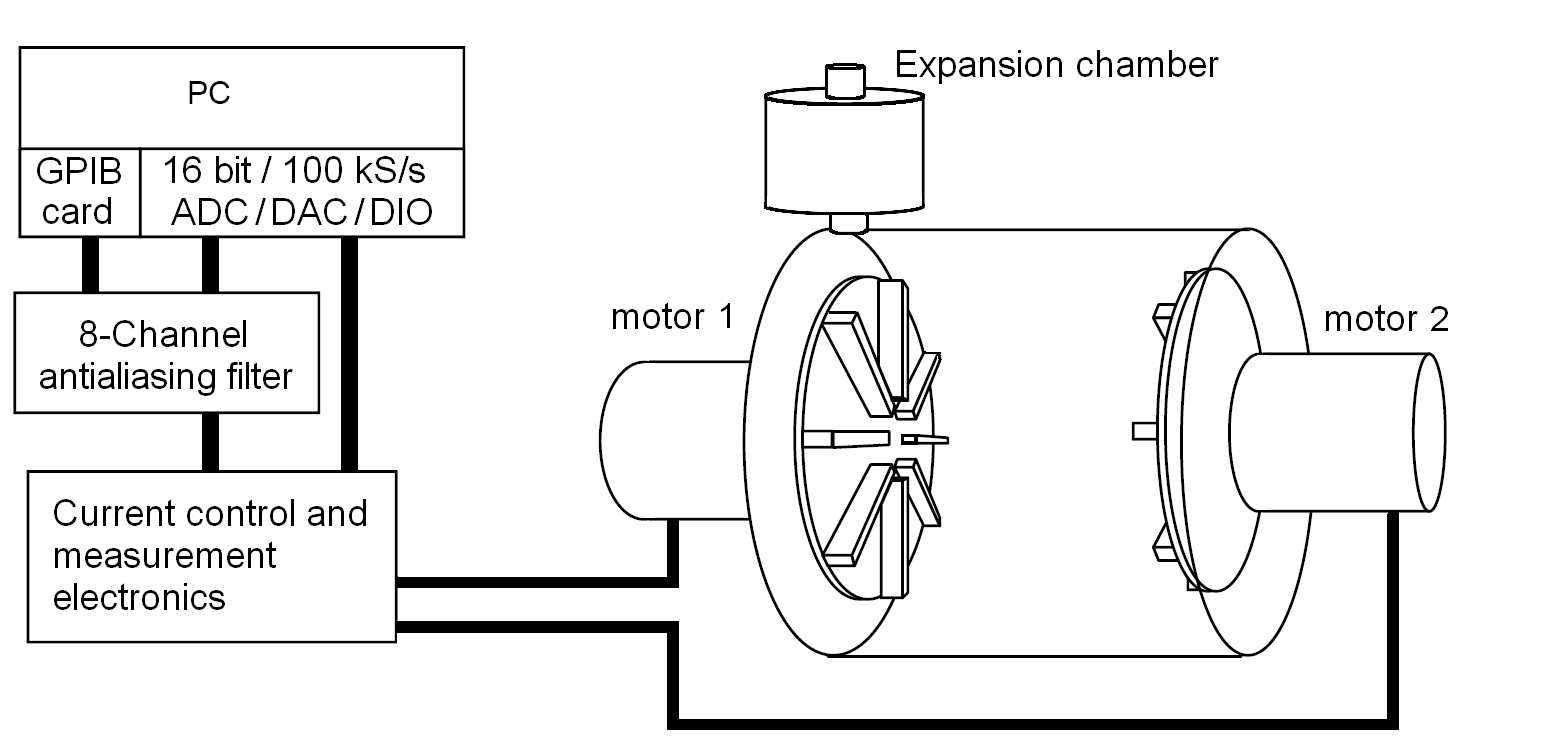}
\caption{A sketch of the experimental setup to produce von K\'arm\'an swirling flows. Driven by electric motors, two coaxial disks with vanes rotate in opposite directions within a cylindrical container filled with a fluid, creating a strong turbulent flow. A PC with a GPIB controller and an ADC/DAC/DIO card controls the experiment and performs the measurements. The current control module delivers current to the electric motors. At the same time, the instantaneous angular speed of each disk is measured. The device sketched here is designed to work with liquids. The small chamber on top provides the room needed for the thermal expansion of the fluid contained in the main volume, and provides a connection to a vacuum pump for degassing purposes.} \label{Fig_1}
\end{figure}

As any dissipative system driven far from equilibrium, turbulent flows require a permanent supply of energy to remain in their non-equilibrium state. In the case of confined von K\'arm\'an swirling flows (see Figure \ref{Fig_1}), the fluctuations in the injected power may have a non-Gaussian statistics, characterized by a probability density function (PDF) strongly asymmetric, with a stretched left side. At least this is the case in experiments in which air is used as the working fluid, and the counter-rotating stirrers are driven at the same \emph{constant} angular speed.\cite{LabPinFau96,PinHolLab99} In addition, it has been shown that the shape of these PDFs remains similar when the Reynolds number of the flow is changed, or at most depends marginally on this parameter in the range where the experiments are typically realized. In a new experiment performed with air, in which the stirrers were driven at constant torque, fluctuations of injected power having a strongly non symmetric PDF were found, this time with the right side stretched towards the high power end.\cite{LabBus12} The reason for this left-right reversal is simple: the constant torque applied by the motors increases the angular speed of the
stirrers when the drag exerted by the flow drops, so that the instantaneous power $P = \Omega\tau$ rises. Contrarily, these events appear as power drops when the speed is held constant. These two types of events have in common the sudden drop in the torque exerted by the flow, and what makes the difference is the stirrers' driving mode. It follows that, in an experiment where the stirrers are driven at constant angular speed, torque and power fluctuations are proportional, and their PDFs are related by
\begin{equation}
\Pi_P (P)dP=\Omega^{-1}\Pi_\tau(\tau)d\tau.
\label{PDF}
\end{equation}
As there is no change in the kinetic energy of the stirrers, this is just the PDF of the power injected \emph{into the flow}. Thus, when the goal of the experiment is to study the statistics of the power injected into the flow, using constant angular speed for the stirrers is the correct choice. Here, and in what follows, we assume $P=\overline{P}+\widetilde{P}$, $\tau=\overline{\tau}+\widetilde{\tau}$, and, of course, $\widetilde{P} = \Omega\widetilde{\tau}$. Over bars and tildes indicate the time average and the fluctuating parts of a quantity, respectively.

When the experiment is performed at constant torque, rises or drops in the injected power increase or decrease both, stirrers and flow kinetic energies.\cite{Note2}   Consequently, the PDF of the total injected power is shaped by both, fluctuations of the power injected into the flow and fluctuations in the kinetic energy of the stirrers. Thus, there is a fundamental difference between an experiment at constant angular speed versus one at constant torque: when a steady regime is reached, in the former there is no net transfer of energy to the stirrers, whereas in the latter the stirrer's changing kinetic energy has a role that cannot be neglected. In an experiment performed by Titon and Cadot\cite{TitCad03} using water as the working fluid, both driving modes were used for the stirrers. Interestingly, they found that the PDF of the injected power is nearly Gaussian at constant torque, and Gaussian at constant angular speed. This last result was confirmed recently by Burnishev and Steinberg\cite{BurSte14} in experiments performed at constant angular speed using pure water and solutions of sugar in water using several concentrations. These results seem to contradict the results obtained in air, because one would expect that in geometrically similar systems the flow must be similar at equal Reynolds numbers. For turbulent flows this statement can be translated into a weaker one: \textit{turbulent flows in geometrically similar systems, and having equal Reynolds numbers, must display similar statistical properties}. Given that the sole parameter of the dimensionless Navier-Stokes equation is the Reynolds number, Re, this seems completely reasonable. Now, Re is related to some characteristic size and speed of the solid boundaries that shape the flow. In von K\'arm\'an flows, the Reynolds number is customarily defined as $\mathrm{Re} = \Omega R^2/\nu$, where $\Omega$ is the angular speed of the stirrers, $R$ is their radius, so that $\Omega R$ is the tangential speed of the disks' edges, and $\nu$ is the kinematic viscosity of the fluid. When $\Omega$ is not tightly controlled, for each stirrer we should expect small fluctuations: $\Omega_i=\overline{\Omega}_i+\widetilde{\Omega}_i$, $i=1,2$. In the symmetric case, we have $\overline{\Omega}_1=\overline{\Omega}_2\equiv \overline{\Omega}$. In addition, when $\widetilde{\Omega}_i=0$, $i=1,2$ we can consider that $\mathrm{Re} = \overline{\Omega} R^2/\nu$ is well defined, even when no much attention is payed to the shape and size of the vanes. A problem arises when a close examination of the meaning of $\Omega = \mathrm{constant}$, in an experimental context, is made. In other words, when a specific experiment is being considered, one could ask two questions:
\begin{itemize}
\item[i)] For small fluctuations in $\Omega_i$, would $\mathrm{Re} = \overline{\Omega} R^2/\nu$  still be a valid definition?
\item[ii)] Is the flow dynamics affected by these fluctuations?
\end{itemize}
In incompressible fluids, including air at low Mach numbers, the flow dynamics is governed by the Navier-Stokes equation. In the symmetric case, with identical and constant angular speeds, the Reynolds number definition given above is perfectly adequate. Now, let us allow small fluctuations in the angular speed of the stirrer 1 by running its driving motor at constant torque, so that $\Omega_1=\overline{\Omega}+\widetilde{\Omega}_1$, while the stirrer 2 rotates in the opposite direction with constant  angular speed: $\Omega_2=\overline{\Omega}$. As in a previous work,\cite{LabBus12} given that the signs of the angular velocities never change, we will work with their magnitudes. Dropping the indices for clarity, we obtain the following governing equations for the flow and stirrer 1 motion:
\begin{subequations}
\label{DynEq}
\begin{equation}
\frac{\partial \mathbf{v}'}{\partial t'}+(\mathbf{v}'\cdot \nabla')\mathbf{v}'=-\nabla' p' + \frac{1}{\mathrm{Re}(\widetilde{\Omega})}\Delta' \mathbf{v}',\label{DynEq:1}
\end{equation}
\begin{equation}
\tau = \int_{S} p\mathbf{r}\times d\mathbf{S} = \rho R^5\overline{\Omega}^{~2}\tau',~~\mathrm{with}~~ \tau' =  \int_{S'} p'\mathbf{r}'\times d\mathbf{S'},\label{DynEq:2}
\end{equation}
\begin{equation}
J \overline{\Omega}\frac{d \Omega}{d t'} +  \gamma_{_\mathrm{M}}\Omega + \rho R^5\overline{\tau}'\Omega^2 = \tau_{_\mathrm{M}} + \widetilde{\tau},
\label{DynEq:3}
\end{equation}
\begin{equation}
t'= \overline{\Omega}t~,~~~~~\mathbf{x}'= \frac{\mathbf{x}}{R}~,~~~~~\mathbf{v}'= \frac{\mathbf{v}}{R\overline{\Omega}}~,~~~~~p'= \frac{p}{\rho R^2\overline{\Omega}^2}~,
\label{DynEq:4}
\end{equation}
\end{subequations}
where $\rho$ is the fluid density, $\gamma_{_\mathrm{M}}$ is the coefficient of ``viscous'' electromagnetic losses in the motor,\cite{Note3} $\tau$ is the torque exerted by the turbulent flow, $\tau_{_\mathrm{M}}$ is the (constant) torque exerted on the armature of the electric motor, $J$ is the stirrer moment of inertia, and $S$ is the union of the leading and trailing surfaces of the vanes. The primed variables in the preceding equations are dimensionless; their relations with dimensional variables are displayed in the equations (\ref{DynEq:4}). The use of dimensional variables in (\ref{DynEq:2}) makes clear the dependence of $\tau$ on the system parameters. The \emph{varying} Reynolds number in equation (\ref{DynEq:1}), defined as $\mathrm{Re}(\widetilde{\Omega})=R^2(\overline{\Omega}+\widetilde{\Omega})/\nu$, is introduced as a simplified mechanism of coupling between stirrer's dynamics and flow dynamics. This premise may be considered as a sort of toy model, and we stress that the results of the analysis that follows do not depend on it. Nevertheless, it effectively closes the loop of the input-output system defined by equations (\ref{DynEq:1}) and (\ref{DynEq:3}). Equation (\ref{DynEq:3}) governs the motion of the stirrer, which is driven by the joint action of the motor torque, $\tau_\mathrm{_M}$, and the flow torque, $\tau$. In this equation, the time is dimensionless, which makes it compatible with the equation (\ref{DynEq:1}). Finally, if the fluctuations are small enough as compared to mean values, and terms up to first order in the fluctuating variables are held, then the equation of motion for the stirrer is reduced to
\begin{subequations}
\label{DynFlu}
\begin{equation}
J \overline{\Omega}\frac{d \widetilde{\Omega}}{d t'} +  \biggl(\gamma_{_\mathrm{M}} + \frac{2\overline{\tau}}{\overline{\Omega}}\biggr)\widetilde{\Omega} = \widetilde{\tau},\label{LangEq:1}
\end{equation}
\begin{equation}
\mathrm{with}~~~\tau = \overline{\tau} + \widetilde{\tau},\label{LangEq:2}
\end{equation}
\begin{equation}
\overline{\tau} = \rho R^5\overline{\Omega}^{~2}\overline{\tau}',~~~\mathrm{and}~~~
\widetilde{\tau} = \rho R^5\overline{\Omega}^{~2}\widetilde{\tau}'.\label{LangEq:3}
\end{equation}
\end{subequations}\\
The solution of the equation (\ref{DynEq:1}), with proper initial and boundary conditions (the stirrer 2 enters here as a moving boundary) provides, through the velocity field, the fluctuating torque $\widetilde{\tau}(t)$ that determines $\widetilde{\Omega}(t)$ in the equation (\ref{LangEq:1}).  Given a velocity field which is a solution of the equation (\ref{DynEq:1}), the dimensionless pressure $p'$ which appears in the equation (\ref{DynEq:2}) can be obtained, in principle, from the Poisson equation for the pressure obtained by taking the divergence of the equation (\ref{DynEq:1}).\cite{Kraich56a} To evaluate the required integrals, a rotating coordinate system attached to the stirrer 1 can be used. Note that this reference frame follows the angular acceleration of the stirrer, so that if we write the equation (\ref{DynEq:1}) in such frame, terms related to Coriolis and Euler forces will appear. Nonetheless, the velocity field seen from any material point should look the same in both, laboratory and rotating reference frames. If $z'_i$ and $u'_i$ are the (dimensionless) coordinates and velocity components in the new reference frame, then the Poisson equation for the pressure is, in its dimensionless form,
\begin{equation}
\frac{{\partial}^2 p'}{\partial {z'_i}^2} = -\frac{\partial (u'_i u'_k)}{\partial z'_i \partial z'_k},\label{Poisson}
\end{equation}
from which the pressure on the surface $S'$ can be calculated. In principle, the problem with constant $\Omega$ can be solved numerically using a specific method to solve (\ref{DynEq:1}) and (\ref{Poisson}). There exist commercial software packages to undertake this problem allowing a number of prescriptions for the eddy viscosity. Not surprisingly, each choice gives a different solution, so that experimental data is required to select the best model. Of course, when $\Omega$ is not merely a constant parameter, but fluctuates as a result of the flow stresses on the stirrer, the preceding approach alone in no longer useful.

From now on, we will use `a' and `w' as subscripts to denote air and water, respectively. In the next two subsections of this introduction, numerical estimates for constant and varying $\Omega$ will be based on parameter values similar to those used in the experiments, namely: $\overline{\Omega}_\mathrm{a}\approx 4 \overline{\Omega}_\mathrm{w}$, $R_\mathrm{a}\approx 2 R_\mathrm{w}$, moment of inertia $J_\mathrm{a}\approx 4.7\times 10^{-3} \mathrm{kg~m^2}$, $J_\mathrm{w}\approx 2\times 10^{-3}\mathrm{kg~m^2}$, and $\mathrm{Re}\approx 1.5\times 10^5$ in both experiments.

\subsection{Constant $\Omega$}

We want to know the effect of a change in the fluid density on some of the system's kinematic and dynamic magnitudes while keeping constant the Reynolds number. Denoting by $\rho_\mathrm{a}$ and $\rho_\mathrm{w}$ the densities of air and water, respectively, and considering the geometries and parameter values used in the experiments, we find that the ratio between the required (constant) angular speeds must be
\begin{equation}
\frac{\Omega_\mathrm{w}}{\Omega_\mathrm{a}} = \frac{\nu_\mathrm{w} R_\mathrm{a}^2}{\nu_\mathrm{a} R_\mathrm{w}^2} = \frac{1}{4}.\label{AngSp}
\end{equation}
From the equation (\ref{LangEq:3}) we have that the relative torque fluctuation is
\begin{equation}
\frac{\tau^\mathrm{rms}}{\overline{\tau}}=\frac{{\tau'}^\mathrm{rms}}{\overline{\tau}'}. \label{TRel}
\end{equation}
As we have the same Re in both cases, the dimensionless velocity field that solves the equation (\ref{DynEq:1}) is the same for air and water, so that the integral giving $\tau'$ in equation (\ref{DynEq:2}) will be also the same in each case. Thus, the ratio in equation (\ref{TRel}) is the same in both cases. Then, the relative torque fluctuation ratio between water and air is
\begin{equation}
\frac{\tau^\mathrm{rms}_\mathrm{w}/\overline{\tau}_\mathrm{w}}{\tau^\mathrm{rms}_\mathrm{a}/\overline{\tau}_\mathrm{a}}=1, \label{TRatio}
\end{equation}
with independence of the ratio $R_\mathrm{w}/R_\mathrm{a}$. Thus, in two given von K\'arm\'an flows having similar geometries, using air in one of them and water in the other, and having the same Reynolds number, the relative rms torque fluctuations produced by the flow are the same. If we consider two systems with similar geometries except by a scale factor, this result will remain valid if the Reynolds number is the same in both devices. Note that the ratio of plain rms torque fluctuations is not independent of some of the systems parameters. This ratio, which of course is equal to the ratio of the mean values, is given by
\begin{equation}
\frac{\tau^\mathrm{rms}_\mathrm{w}}{\tau^\mathrm{rms}_\mathrm{a}}=\frac{\overline{\tau}_\mathrm{w}}{\overline{\tau}_\mathrm{a}} = \frac{\rho_\mathrm{w}\nu_\mathrm{w}^2 R_\mathrm{w}}{\rho_\mathrm{a}\nu_\mathrm{a}^2 R_\mathrm{a}}\approx 1.4, \label{TRatio_1}
\end{equation}
where the numerical value is what would result from an experiment done at constant angular speeds.\\

The previous results are a direct consequence of the principle of dynamic similarity, which implies the condition of constant angular speed for the stirrers. Equivalently, these are the implications of assuming that the fluctuating part of the hydrodynamic forces have no effect on the stirrers' motion.

\subsection{Constant $\tau$}

The problem becomes far less simple when $\Omega$ fluctuates, because in this case the hydrodynamics must be coupled, through the solution of equation (\ref{Poisson}) and (\ref{DynEq:2}), to the motion equation (\ref{DynEq:3}) of the stirrer. In addition, we expect changes in the flow structure when the vanes perform an accelerated motion in response to the flow action. In fact, the equations (\ref{DynEq:1}) and (\ref{DynEq:3}) conform a closed-loop dynamical system with parametric feedback: the input of the equation (\ref{DynEq:3}) is a functional of the velocity field that solves the equation (\ref{DynEq:1}), and the latter is parametrically coupled to the output of the former through the coefficient of the Laplacian term. This means that the effect of $\widetilde{\Omega}$ upon the velocity field depends in a complicated manner on the value of $\Omega$ ($=\overline{\Omega}+\widetilde{\Omega}$), and its time derivatives. Inserted in the loop, we have the flow with its own dynamics, which we can attempt to understand through its effects on the stirrers motion. Although this problem cannot be easily solved by numerical methods, the rather obvious role of the fluid density is made clear by the equation (\ref{DynEq:2}): the torque $\tau$ exerted by the flow, and its fluctuating part, $\widetilde{\tau}$, are proportional to the density $\rho$, so that if we consider a system with constant $\Omega$ in which we only replace the air with water, the rms amplitude of torque fluctuations must change by a factor equal to the ratio of the densities: $\rho_\mathrm{w}/\rho_\mathrm{a}\approx 830$.\\

For the device filled with water, the stirrer's motion equation in dimensional variables is
\begin{equation}
J \dot{\widetilde{\Omega}}_\mathrm{w} +  \biggl(\gamma_{_\mathrm{M}} + \frac{2\overline{\tau}_\mathrm{w}}{\overline{\Omega}_\mathrm{w}}\biggr)\widetilde{\Omega}_\mathrm{w} = \widetilde{\tau}_\mathrm{w}.\label{MEqW}
\end{equation}
In the limit of vanishing moment of inertia $J$, the ratio between the rms amplitudes of angular speed and torque is simply
\begin{equation}
\frac{\widetilde{\Omega}_\mathrm{w}^\mathrm{rms}}{\widetilde{\tau}_\mathrm{w}^\mathrm{~rms}} = \frac{\overline{\Omega}_\mathrm{w}}{\gamma_{_\mathrm{M}}\overline{\Omega}_\mathrm{w} + 2\overline{\tau}_\mathrm{w}}.\label{ROT}
\end{equation}
At this point, it is necessary to assume that the weak similarity principle stated before is valid in this context. If the Reynold number ($\mathrm{Re}=R^2\overline{\Omega}/\nu$) has the same value for the flows using water and air, then their statistical properties should be similar. Therefore, for fluctuations in a very low frequency band or, equivalently, for a vanishing moment of inertia, the ratio of the preceding fraction between air and water should be
\begin{equation}
\frac{\widetilde{\Omega}_\mathrm{a}^\mathrm{rms}/\widetilde{\tau}_\mathrm{a}^\mathrm{~rms}}{\widetilde{\Omega}_\mathrm{w}^\mathrm{rms}/\widetilde{\tau}_\mathrm{w}^\mathrm{~rms}} \approx \frac{\overline{\Omega}_\mathrm{a}\overline{\tau}_\mathrm{w}}{\overline{\Omega}_\mathrm{w}\overline{\tau}_\mathrm{a}} = \frac{\rho_\mathrm{w} R_\mathrm{w}^5 \overline{\Omega}_\mathrm{w}}{\rho_\mathrm{a} R_\mathrm{a}^5 \overline{\Omega}_\mathrm{a}}\approx 5.4,\label{RWA}
\end{equation}
where we neglected the motor losses. At higher frequencies the moment of inertia of the stirrers becomes important, because of the increasing loss of coherence between the stirrer rotation and the spatially averaged rotation of the flow. Still neglecting the motor losses, the equation (\ref{MEqW}) implies that there is a cutoff frequency $f_\mathrm{c}$ for the angular speed fluctuations given, in general, by
\begin{equation}
f_\mathrm{c} \approx \frac{2\overline{\tau}}{J\overline{\Omega}} = \frac{2\rho R^5 \overline{\Omega}\overline{\tau}'}{J}.\label{fcut}
\end{equation}
Now, for two devices running at equal Reynolds numbers, one with water and the other with air, we obtain the following ratio for the cutoff frequencies:
\begin{equation}
\frac{f_\mathrm{c}^\mathrm{w}}{f_\mathrm{c}^\mathrm{a}} \approx \frac{\rho_\mathrm{w} J_\mathrm{a} R_\mathrm{w}^5 \overline{\Omega}_\mathrm{w}}{\rho_\mathrm{a} J_\mathrm{w} R_\mathrm{a}^5 \overline{\Omega}_\mathrm{a}}\approx 13,\label{fcdm}
\end{equation}
where the parameter values are those used in our experiments. For the dimensionless cutoff frequencies, we obtain the ratio
\begin{equation}
\frac{f_\mathrm{c}^\mathrm{w}/\overline{\Omega}_\mathrm{w}}{f_\mathrm{c}^\mathrm{a}/\overline{\Omega}_\mathrm{a}} \approx 51.\label{fcdml}
\end{equation}
These numerical values are a direct consequence of assuming that the weak similarity principle ---valid when $\Omega=\mathrm{constant}$--- can be extended to systems where the characteristic velocity have small fluctuations. Of course, the same kind of analysis can be carried out when some geometric parameter undergoes small fluctuations, or even when some of the fluid parameters, like density or viscosity, undergoes global fluctuations of small amplitude.\\

Given its overall complexity and implications, it seems worth to design an experiment to gather data allowing some further understanding on this subject. It would provide some specific results to compare with the estimates obtained above, and possibly some insight on the way in which the energy injected by the stirrers is transferred to the flow. In what follows, we describe the experimental setup in Section \ref{Setup}, and give the results of the spectral and statistical analysis, which will show that there are substantial differences between air and water statistics. Next, in Section \ref{ETD} the results of the cross correlation study of the energy transfer dynamics are reported. They will make clear that the system dynamics, as well as the energy transfer dynamics, are markedly different when the working fluid is water instead air. In Section \ref{Conc} we compare the scaling of some additional dynamic magnitudes with the experimental results and draw our conclusions. In the Appendix A we give details about the signal processing used in the experiment with water. Finally, in the Appendix B we develop a simplified analysis about the dynamics of experiments performed at constant torque vs constant angular speed.

\section{Experimental setup and measurement results}
\label{Setup}

To answer the questions i) and ii) in the previous section, an experiment was designed in which two geometrically similar devices running at constant torque produce von K\'arm\'an swirling flows at nearly the same Reynolds number: in one of them the working fluid is air while in the other it is water. In each device the power injected to the system can be derived simply from the product between the sum of the measured angular speed of the stirrers and the torque $\tau$ applied to them by the electric motors, which in this case is held constant:\cite{Note4}
\begin{equation}
P(t) = \tau[\Omega_1(t) + \Omega_2(t)] = 2\tau \overline{\Omega} + \tau[\widetilde{\Omega}_1(t)+\widetilde{\Omega}_2(t)]=\overline{P}+\widetilde{P}(t).
\label{Pow1}
\end{equation}
Here, we will be focused in the results of the measurements of $\widetilde{P}_1(t)$, $\widetilde{P}_2(t)$ and $\widetilde{P}(t)=\widetilde{P}_1(t)+\widetilde{P}_2(t)$, in water and air.

In air, the data were obtained using the experimental setup described in a previous work.\cite{LabBus12} For the experiment in water, the apparatus sketched in Fig. \ref{Fig_1} was designed and built. It is basically a half-scale version of the system used with air. Let us assume that the mean values are not strongly affected by the fluctuations discussed in the previous section. By using dimensional analysis, it is easy to estimate the power required by this device. What we want is a device having a power consumption of the same order of magnitude than that of the device for experiments with air. Therefore, we need to calculate a geometric scale factor $\lambda$, which applied to the dimensions of the device used with air gives the right lengths and ratios of the device for water. Given that both Reynolds numbers must be equal, we have
\begin{equation}
\mathrm{Re} = \frac{\overline{\Omega}_\mathrm{w} R_\mathrm{w}^2}{\nu_\mathrm{w}} = \frac{\overline{\Omega}_\mathrm{a} R_\mathrm{a}^2}{\nu_\mathrm{a}},
\label{Re_aw}
\end{equation}
so that the radii are related by
\begin{equation}
R_\mathrm{w} = \sqrt{ \frac{\nu_\mathrm{w}\overline{\Omega}_\mathrm{a}}{\nu_\mathrm{a}\overline{\Omega}_\mathrm{w}}}R_\mathrm{a}.
\label{Rratio}
\end{equation}
 In experiments performed with water or air, we typically find that the angular speed ratio between water and air is $\overline{\Omega}_\mathrm{w}/\overline{\Omega}_\mathrm{a}=1/4$, so that the scale factor between the devices for water and air is
\begin{equation}
\lambda = \frac{R_\mathrm{w}}{R_\mathrm{a}}\approx 0.52.
\label{FScale}
\end{equation}
With this scale factor, the ratio between the mean power consumption in water and air is
\begin{equation}
\frac{\overline{P}_\mathrm{w}}{\overline{P}_\mathrm{a}} = \frac{\rho_\mathrm{w}}{\rho_\mathrm{a}} \biggl(\frac{R_\mathrm{w}}{R_\mathrm{a}}\biggr)^5\biggl(\frac{\overline{\Omega}_\mathrm{w}}{\overline{\Omega}_\mathrm{a}}\biggr)^3\approx 0.48,
\label{FScale_1}
\end{equation}
while the ratio between the mean torques is
\begin{equation}
\frac{\overline{\tau}_\mathrm{w}}{\overline{\tau}_\mathrm{a}} = \frac{\rho_\mathrm{w}}{\rho_\mathrm{a}} \biggl(\frac{R_\mathrm{w}}{R_\mathrm{a}}\biggr)^5\biggl(\frac{\overline{\Omega}_\mathrm{w}}{\overline{\Omega}_\mathrm{a}}\biggr)^2\approx 1.5.
\label{FScale_2}
\end{equation}
\begin{figure}[t]
\centering \vspace{-0.5cm} \hspace{-0.2 cm}
\includegraphics[width=1.00\textwidth]{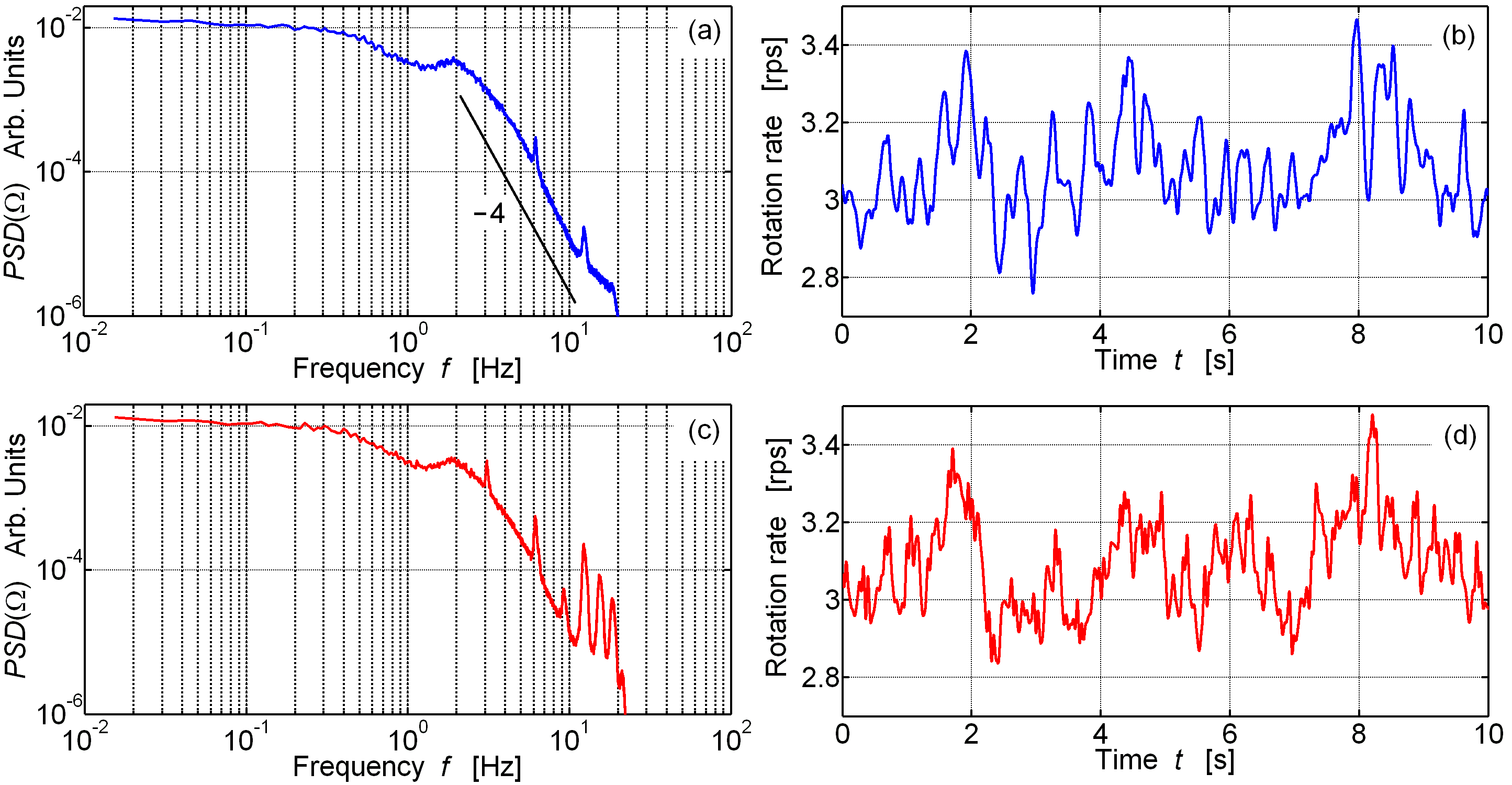}
\caption{(Color online) Rotation rate fluctuations of the stirrers in the device shown in Figure~\ref{Fig_1}. In the subplot (a) and (c), spectra showing roll-off regions with slope $s=-4$ can be seen. In the frequency range from $0.4$~Hz to $1$~Hz a less pronounced roll-off of the spectra can be seen, corresponding to a first order, Langevin-like dynamics (see text). Low pass filters with cutoff frequency $f_c=19$~Hz were used to remove noise (see text). In subplots (b) and (d), short records of the corresponding rotation rate signal are displayed. The curve in subplot (d) looks somewhat more noisy than the curve (b), which is reflected in the spectrum (c) by three peaks close to the cutoff. Note that a degree of correlation exists between the curves in plots (b) and (d).} \label{Fig_2}
\end{figure}
Then, when both experiments run at equal Reynolds numbers, the experiment with water requires 50\% more torque than the required with air, and nearly half the power. In our case, the height of the vanes in the water device was reduced by slightly more than $50\%$, so that the resulting power consumption is about $34\%$ of that required by the device working with air. We remark that a small reduction in the height $h$ of the vanes has no noticeable effect on the statistical properties of the power fluctuations. In fact, $h$ controls the pumping action of the stirrers. Roughly speaking, the radial pumping is related to the volume of fluid contained between the vanes, and this volume scales linearly with $h$. Thus, changing $h$ changes proportionally the radial mass flow rate. We illustrate this point at the end of this section with a measurement in air of the effect that a substantial change in the geometry of the vanes has on the average injected power and its fluctuations.\\

To obtain similar Reynolds numbers in both devices, the angular speed of the stirrers in air must be approximately four times greater than the angular speed in water. Taking this into account, for measuring angular speeds in water we used optical encoders with twice the resolution of those used for air. This allowed high quality measurement of fluctuations at very low rotation rates. Demineralized water was used, and to avoid bubbles it was degassed by using a rotary vane vacuum pump. After degassing, no bubbles were visible within the apparatus running either in co-rotating or counter-rotating modes. Data acquisition and processing were performed as described in a previous work \cite{LabBus12}. The stirrers' mean rotation rates, $f=\Omega/2\pi$, were ${\overline f}_\mathrm{a} \approx 11.4$~rps (revolutions per second) with air and ${\overline f}_\mathrm{w} \approx 3.1$~rps with water, giving Reynolds numbers $Re_\mathrm{a} \approx 1.55\times 10^5$ and $Re_\mathrm{w} \approx 1.58\times 10^5$ with air and water, respectively. In both experiments the signals were low-pass filtered, using cutoff frequencies $f_c^\mathrm{a} = 20$~Hz and $f_c^\mathrm{w} = 19$~Hz. This latter cutoff is necessary because the asymmetries and noise of the electric motors used with water produced angular speed fluctuations with an amplitude of about $38\%$ of the amplitude of the fluctuations produced by the turbulent flow. It is important to stress that the only effect of filtering on the PDFs obtained with water is a slight reduction in its width. No noticeable change in their shape after the filtering process is observed. In air, where pancake DC servomotors were used, the noise and the asymmetries of the motors are small, but filtering improves the calculation of the angular acceleration from the angular speed data. A detailed explanation of the signal processing used for the experiment with water is given in Appendix \ref{Ap_A}.\\

Figure \ref{Fig_2} displays spectra and signals corresponding to the rotation rate fluctuations of the stirrers for the experiment with water. The upper and lower spectra displayed on the left side, corresponding to the left and right stirrers, respectively, have three clearly different zones: i) a flat region in the lowest frequency band, spanning a little more than one decade, ii) a short decay between $0.4$~Hz and $\gtrapprox 1$~Hz, and iii) a roll-off region with a scaling $\sim f^{-4}$ for $f>2$~Hz. This latter zone is the combined result of the continued $f^{-2}$ roll-off starting in ii) plus an additional $f^{-2}$ roll-off, possibly related to an averaging process of normal stresses, on the surface of the vanes, related to flow structures whose characteristic length goes from the height of the vanes down to the Kolmogorov scale.

The resulting angular speed signals are similar to those obtained in air. Although one of the stirrer has some increased noise below the cutoff frequency of the low-pass filter, its amplitude is too small to have a significant effect on the signal statistics.
\begin{figure}[t]
\centering \vspace{-0.5cm} \hspace{-0.2 cm}
\includegraphics[width=1.00\textwidth]{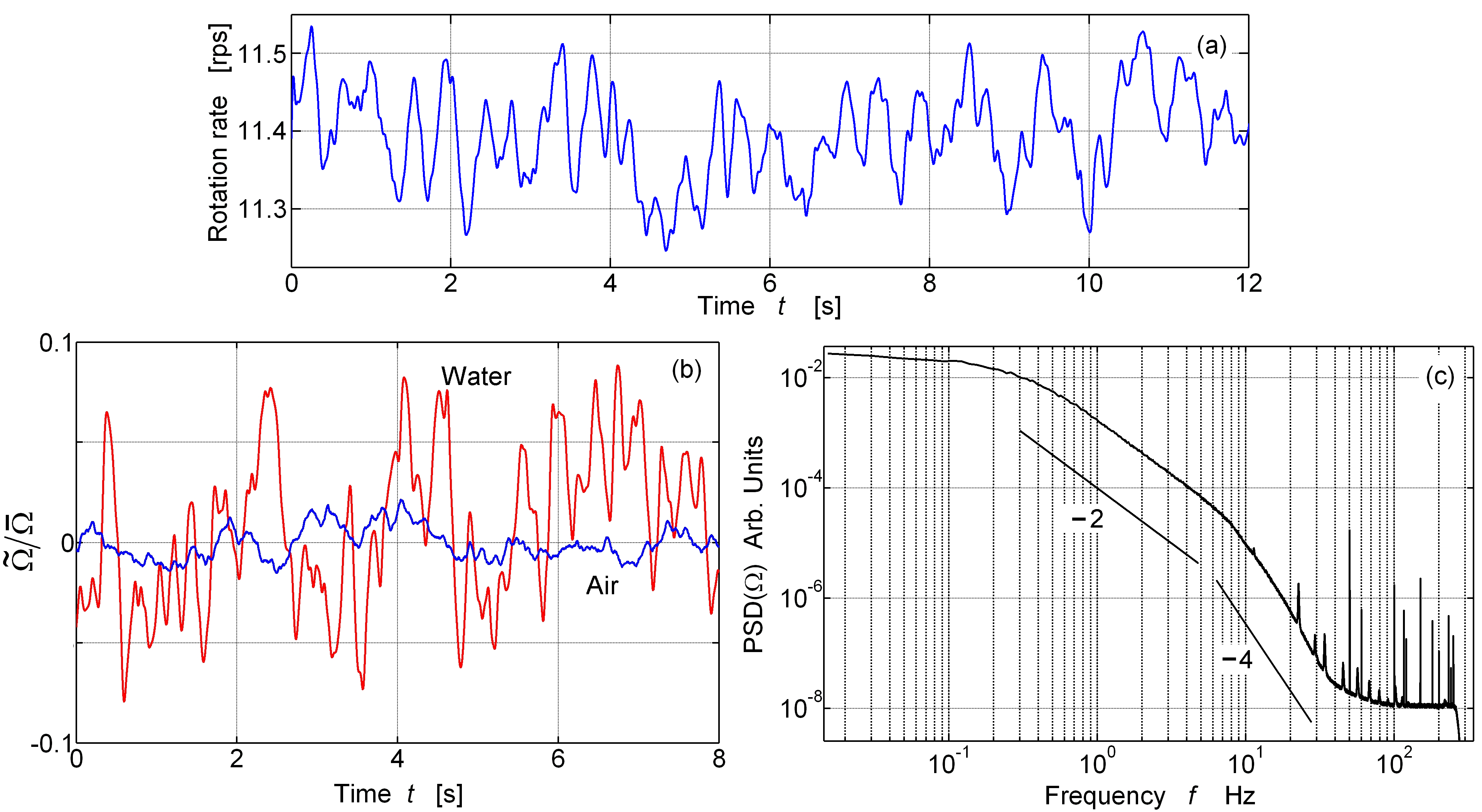}
\caption{(Color online) (a) Rotation rate of one of the stirrers for the experiment in air. (b) Fluctuations of the rotation rate divided by the mean rotation rate in water (red) and air (blue). Note that the relative amplitude is much higher in water, due to the effect that a density about $830$ times larger has. (c) Spectrum of rotation rate fluctuations for the experiment in air. In this case two distinct regions, with slope $-2$ and $-4$, exist (see text).} \label{Fig_3}
\end{figure}

\begin{figure}[t]
\centering \vspace{-0.5cm} \hspace{-0.2 cm}
\includegraphics[width=1.00\textwidth]{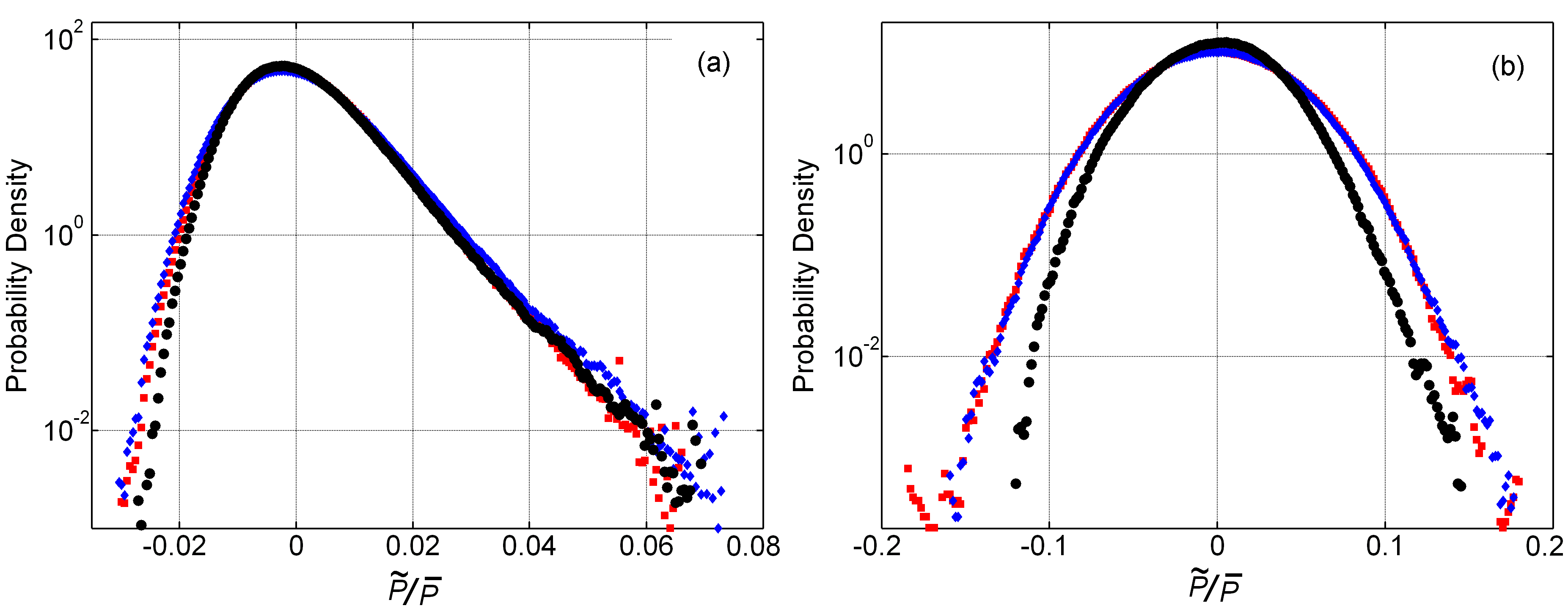}
\caption{(Color online) PDF of power fluctuations divided by the mean power in air (a) and water (b). Stirrer 1 is represented by red squares and stirrer 2 by blue diamonds. Black circles represent the total injected power (TIP). Note that the relative amplitude of fluctuations is much higher in water, due to its greater density. It is clear from the plot (b) that in water the PDF of injected power is \textit{nearly} Gaussian (see text).} \label{Fig_4}
\end{figure}

On the other hand, the experiment performed in air produces signals like the one displayed in Figure \ref{Fig_3}(a). The relative amplitude of this signal, $\widetilde{\Omega}/\overline{\Omega}$, is compared with the corresponding signal obtained in water in Figure \ref{Fig_3}(b). We can see that in water the relative amplitude of the angular speed fluctuations is about five times larger than in air. Figure \ref{Fig_3}(c) displays the raw spectrum of the signals corresponding to angular speed fluctuations obtained in air. It can be seen that this spectrum is qualitatively different to the spectra displayed in Figure \ref{Fig_2}(a) or (b): there is a wide region with a roll-off that scales as $f^{-2}$, which is barely present in the spectra obtained in water. We will return to this point later.  Given that in each experiment the torque $\tau$ applied to each stirrers is in principle the same, the total injected power (TIP) can be calculated in both cases using equation (\ref{Pow1}).
The PDFs of $\widetilde{P}(t)/\overline{P}$ obtained in both experiments are displayed in Figure \ref{Fig_4}. The huge difference between these results is apparent. In (a), the PDFs obtained with air have the same shape than those obtained previously on this device,\cite{LabBus12} as expected. In (b), the PDFs obtained in water are \textit{almost} Gaussian, a result closer to those reported in former works.\cite{TitCad03,BurSte14} A possible explanation for the difference between air and water begins to arise when we look at the spectra of these fluctuations. In air ---Figure \ref{Fig_3}~(c)--- the spectrum is characterized by the presence of three regions: i) a nearly flat zone, below $0.15$~Hz, ii) a first roll-off scaling as $f^{-2}$ for almost one and half decade, followed by iii) a second roll-off with scaling $f^{-4}$. In a recent work using air,\cite{LabBus12} it was shown that the dynamics of $\widetilde{\Omega}(t)$ on regions i) and ii) is governed by a Langevin equation, obtained by linearizing the equations of motion. Deconvolution of the  the angular speed signals revealed that the fluctuations of the torque  exerted by the flow on the stirrers have a flat spectrum, at least in the range of frequencies below the end of region ii). This coincides with the spectrum of torque fluctuations measured in air at constant angular speed. Probably a flat spectrum still holds at higher frequencies, which implies ---somewhat surprisingly--- that the spectrum of torque fluctuations resembles that of a white noise, despite the fact that it comes from the integral of normal stresses on the surface of the vanes ---a sort of weighted sum. The roll-off in region iii) cannot be explained in terms of the stirrer's mechanical response. The interpretation for this behavior is the following: the frequencies belonging to the spectrum zone that scales as $f^{-4}$ are related to flow scales that become comparable or smaller than the height of the vanes,\cite{LabBus12} so that their contribution to the total torque adds up increasingly incoherently to the surface integral at smaller scales or, equivalently, higher frequencies. The effect on the spectrum is an additional $f^{-2}$ roll-off, which combined with the $f^{-2}$ fall related to the stirrers inertia, gives the $f^{-4}$ region. In water ---Figure \ref{Fig_2}~(a)--- the spectrum still has three regions, but the intermediate region, corresponding to region ii) in the spectrum for air, is nearly nonexistent: the flat, low frequency region, extends up to about $0.5$~Hz, then the (collapsed) middle region goes up to $2$~Hz, and finally we see the region with roll-off $f^{-4}$ up to the sharp cutoff of the noise filter. Thus, the spectrum observed in water is qualitatively different from the spectrum obtained with air: the shapes are clearly different. This implies that the time-domain dynamics of these two systems is not the same. If the dynamics of each system is different, we cannot invoke the similarity principle, not even the weak version given on the first paragraph, to state that the power fluctuations in water and air should be similar. In these experiments, using geometrically similar devices and similar Reynolds numbers, that is, in conditions where the hydrodynamic similarity principle holds, different results are obtained when the fluid is water instead air. Being this the case, it is not surprising that experiments performed in water\cite{TitCad03,BurSte14} give results different from those obtained in air\cite{LabPinFau96,PinHolLab99,LabBus12}.
Now, if we look at the PDFs, we note that the amplitude of relative fluctuations for the TIP, $\widetilde{P}/\overline{P}$, is \emph{smaller} than the relative amplitude of the individual stirrers. In Figure \ref{Fig_4}, the PDF of the total power is represented by black circles. In (a), we see that this effect is rather small, whereas in (b) it is clearly visible. This marks another difference in the dynamics when water is used instead air: fluctuations in the rotation speed have an anticorrelated component which in water is stronger than in air. This anticorrelation characterizes the global rotation of the flow, a behavior that has its own dynamics and scaling properties, as shown in a previous experiment in air,\cite{LabBus12} and belongs to the motion dynamics on the lowest frequency range of the spectrum.

\begin{figure}[t]
\centering \vspace{-0.5cm} \hspace{-0.2 cm}
\includegraphics[width=1.00\textwidth]{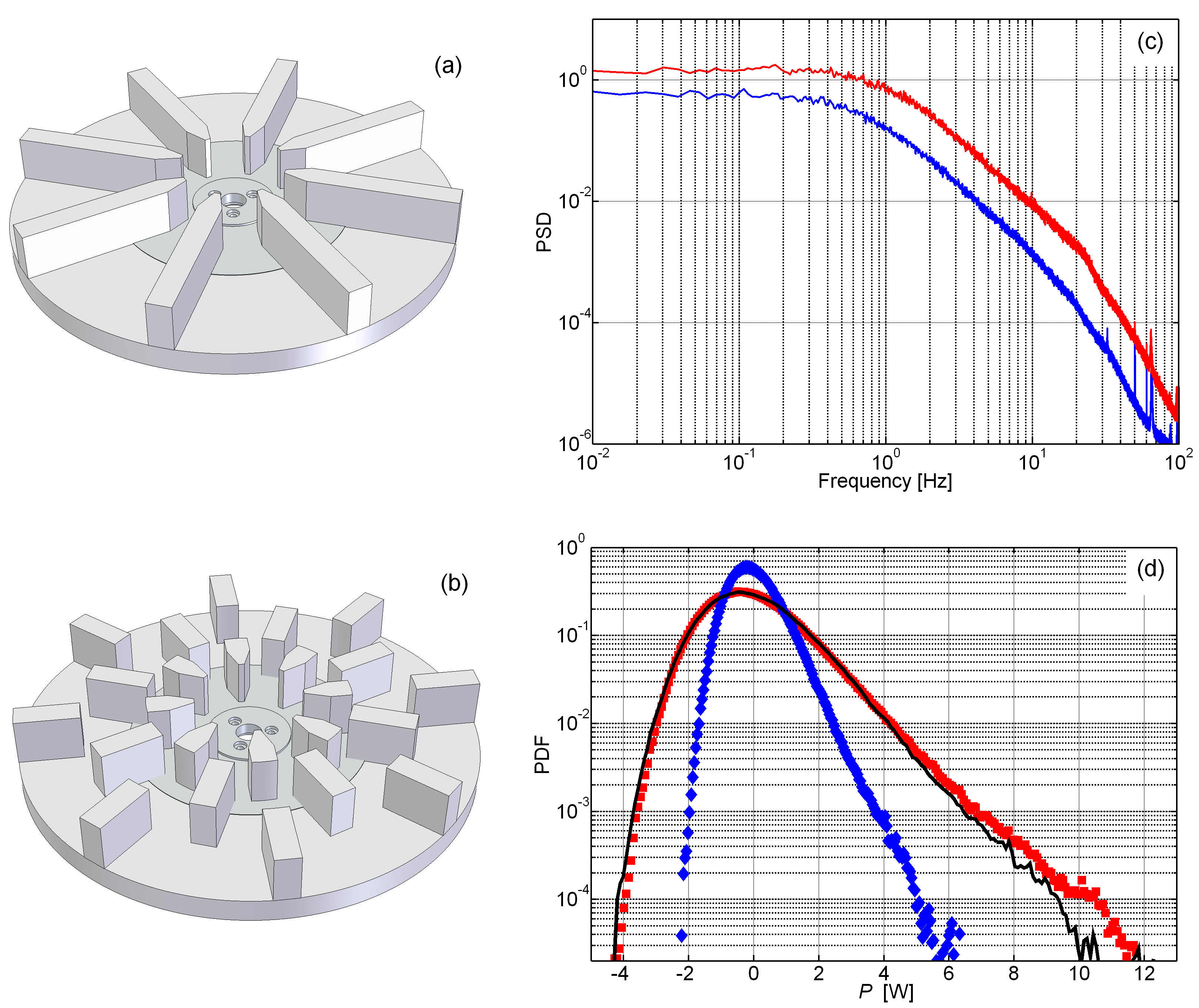}
\caption{(Color online) Effect of vane geometry on the global statistics. (a) Disk with continuous vanes, used for the experiment with air. (b) Disk with segmented vanes. It has the same size and the same moment of inertia that the disk displayed in (a). (c) Spectra of total injected power. The lower curve has an overall reduced energy, because of the reduction of the total power injected to the flow when diks with segmented vanes are used. Note that the shape of the curves remains essentially unchanged. (d) PDFs of the injected power, corresponding to the experiment with continuous vanes (red squares) and the measurement using segmented vanes (blue diamonds). The PDF corresponding to the latter case was scaled (black, continuous curve) to show that its shape undergoes only minor changes, as compared to the PDF corresponding to continuous vanes.} \label{Fig_5}
\end{figure}

As mentioned earlier, it may be shown that a major change in the shape of vanes has a minor effect on the PDF and the spectrum of the injected power. The present test was made with air, but by using water one should extract similar conclusions. Figure \ref{Fig_5}~(a) displays the disk with the vanes used in the experiment with air. Figure \ref{Fig_5}~(b) displays a disk with segmented vanes. The discontinuities in the vanes greatly affect the radial mass flux. This results in a reduction of the power required to maintain a given mean angular speed $\Omega$. Specifically, disks with continuous vanes require a total mean power of $\overline{P}_\mathrm{c}\approx 900$~W to maintain a rotation rate $\overline{\Omega}/(2\pi)=32$~rps, whereas with segmented vanes, it is enough with $\overline{P}_\mathrm{s}\approx 690$~W.  This makes a reduction of $23$\% in the injected power. Figure \ref{Fig_5}~(c) displays the normalized spectra of injected power in both cases. As can be seen, the change in the shape is marginal. The upper curve (red) was obtained with continuous vanes. There is a small reduction in the first cutoff frequency of the lower (blue) curve, and the ratio between this one and the second cutoff is slightly smaller, as compared with the upper curve.

The wide curve (red squares) in Figure \ref{Fig_5}~(d) is the PDF of the TIP obtained with continuous vanes, whereas the narrow curve (blue diamons) is the PDF of the TIP issued from disks with segmented vanes. It has approximately half the width of the former curve, that is, the ratio between the widths is  $s=\sigma_\mathrm{c}/\sigma_\mathrm{s}\approx 2$. However, when the height of the later PDF is divided by $s$, and its width is multiplied by $s$, the result is the continuous curve (black) which is almost coincident with the PDF obtained with continuous vanes, as shown in Figure \ref{Fig_5}~(d). Thus, this result shows that the statistics of the injected power is fairly insensitive to the shape of the vanes. Note that the symmetry of the stirrers with continuous vanes belongs to the D$_8$ symmetry group. This symmetry implies that, for stirrers rotating in opposite directions with equal, constant angular speed, the flow has the following symmetry property under simultaneous rotation reversal of the stirrers:
\begin{equation}
\biggl\{{ \mathbf{\Omega}_1 \atop   \mathbf{\Omega}_2 }\biggr\} \rightarrow \biggl\{{-\mathbf{\Omega}_1 \atop -\mathbf{\Omega}_2 }\biggr\} \Rightarrow  v_\phi \rightarrow -v_\phi,
\label{RSymm}
\end{equation}
where $v_\phi$ is the azimuthal component of the flow velocity. This property is preserved by the disks with segmented vanes, which implies that by reversing the rotation of both disk, the only effect on the flow will be $v_{\phi}\rightarrow-v_{\phi}$, in both cases. Previous works give a clue about the effect of modifying the vanes without preserving the stirrers' primary symmetry. In the experiment reported by Burnishev and Steinberg, curved vanes were used. The corresponding symmetry group is C$_8$, so that in their experimental device the symmetry property (\ref{RSymm}) is lost. Although we might not necessarily expect a change in the statistics of power fluctuations, we should expect a noticeable change in the mean flow. In fact, the radial pumping by the vanes is enhanced,  and less angular momentum is injected to the flow by each stirrer. This change in the geometry has been used in experiments to study the dynamo action in von K\'arm\'an swirling flows, using melted sodium as working fluid, in order to obtain similar poloidal and toroidal velocities.\cite{BouMarPet02} Despite the loss of the symmetry property (\ref{RSymm}) and the change in the ratio between poloidal and toroidal components of the motion, we see that Burnishev and Steinberg still find a Gaussian statistics for the injected power, as in the experiments with straight vanes performed by Titon and Cadot. This suggest that changing the shape of the vanes, even if a loss in the system's symmetry is involved, has only a marginal effect ---if any--- on the statistics of the injected power. However, with regard to the global flow, some experiments have shown that stirrers with curved vanes can produce inverse turbulent cascades and a bistable dynamics in the global flow.\cite{LopBur13} A detailed study of different regimes and the supercritical transition to fully developed turbulence, in a von K\'arm\'an flow driven by stirrers with curved vanes, can be found in the work by Ravelet \textit{et al.}\cite{RavChiDav08}

\section{Energy transfer dynamics}
\label{ETD}

In the previous section we have seen that the PDFs of $\widetilde{\Omega}$ obtained in air and water are markedly different. Given that the fluctuations are finally due to the flow in a neighborhood of the stirrers, the observed difference in the PDFs is a strong indication of differences in the flow itself, although the PDFs do not have explicit information about the flow dynamics, nor its energy transfer dynamics. We can undertake this last aspect by looking at time cross correlations of the power components involved in the energy transfer.

\begin{figure}[t]
\centering \vspace{-0.5cm} \hspace{-0.2 cm}
\includegraphics[width=1.00\textwidth]{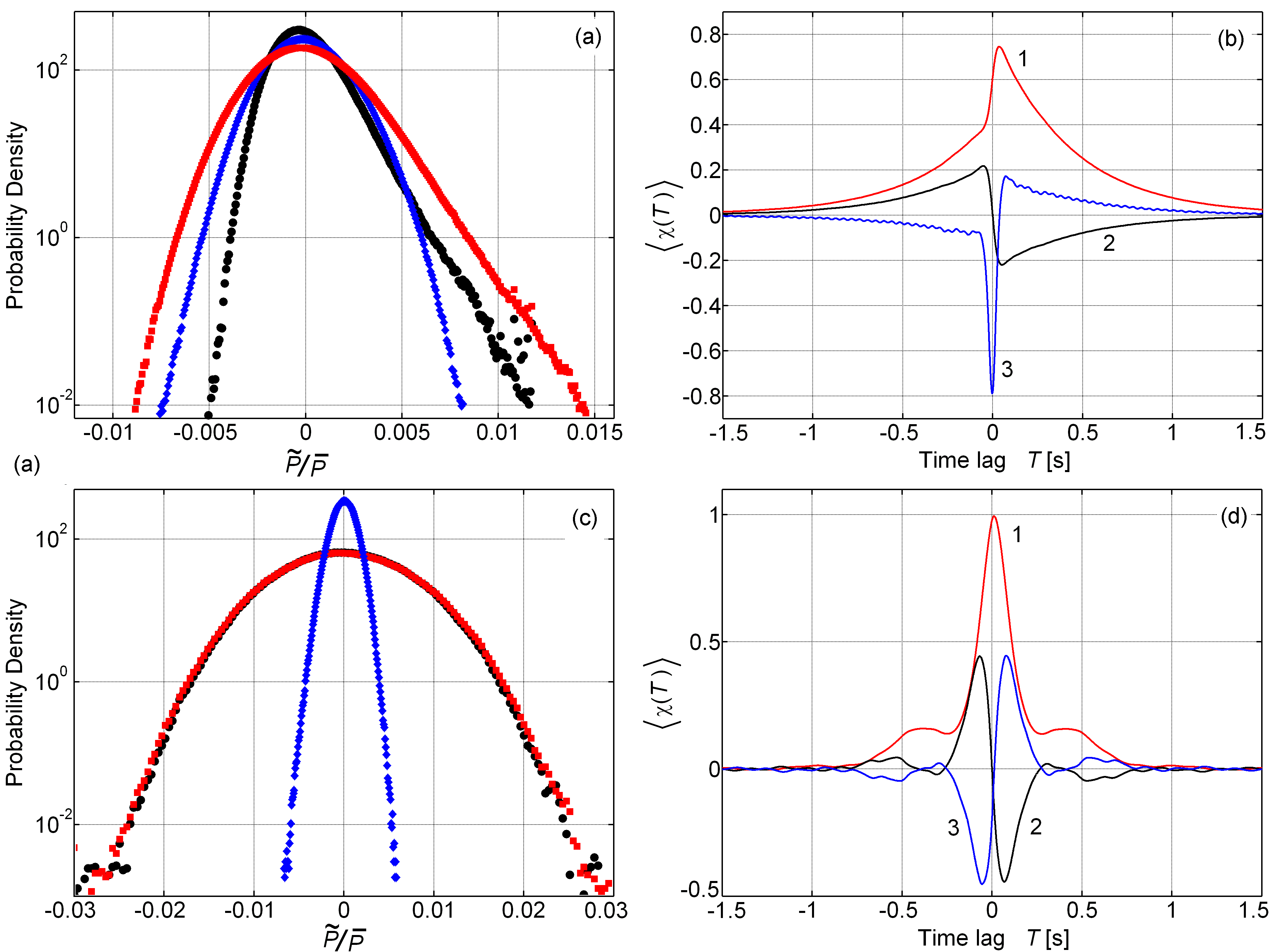}
\caption{(Color online) (a) PDFs of fluctuations---in air, normalized to the mean total power--- of total injected power (TIP) (black circles), Stirrer power (blue diamonds), and power transferred to the Flow (red squares). (b) Mean cross-correlation functions between fluctuating powers: Flow-TIP (1, red); Stirrer-TIP (2, black); and Flow-Stirrer (3, blue). The time lag between Flow power fluctuations and TIP is $T_a = 36$~ms. In water (c), the PDFs of injected power (black, circles) and  the power transferred to the flow (red, squares) are basically coincident. Note that the black (circles) and red (squares) PDFs are about twice as wide as those in (a), whereas the stirrer power PDF (blue, diamonds) is narrower. The mean cross-correlation functions (d) also differ remarkably from those obtained in air. In water, the time lag between fluctuations of Flow power and TIP is $T_\mathrm{w}=11.3$~ms (see text).} \label{Fig_6}
\end{figure}

Let us take a closer look at one stirrer in air. After some straightforward algebra, it can be shown that the total power fluctuation is related to the sum of the power delivered to the stirrer and the flow by
\begin{equation}
\widetilde{P} = J(\overline{\Omega} + \widetilde{\Omega})\dot{\Omega} +  \widetilde{P}_f
\label{Split_P}
\end{equation}
where $\widetilde{P}$ is the fluctuation of the TIP, $J$ is the stirrer's moment of inertia (which includes the motor armature), and $\widetilde{P}_f$ is the fluctuation of the power injected to the flow. The first term on the right hand side is the stirrer power consumption, which indeed represent a `reactive' component, because it does not dissipate energy. In addition, the main contribution of this term to $\widetilde{P}$ comes from $J\overline{\Omega}\dot{\Omega}$, because $J\widetilde{\Omega}\dot{\Omega}$ is about one hundred times smaller. Figure \ref{Fig_6}~(a) displays the PDFs of these three quantities. We see that the PDFs of both, the TIP and the power transferred to the flow are strongly asymmetric. The reactive power spent in accelerating the stirrer seems nearly Gaussian, but is still asymmetric, with positive skewness. In Figure \ref{Fig_6}~(b) the mean cross-correlation functions between these magnitudes can be seen. The curve 1 (red) shows the cross-correlation between the flow power consumption and the TIP. The retarding action of the stirrer on the energy flow is evidenced by the time lag
\begin{equation}
T_\mathrm{a}=36~\mathrm{ms}\label{TlagAir}
\end{equation}
of its peak. Thus, the stirrer operates as a momentary energy storage, as can be also deduced from the anti-correlation dip at zero time lag in the curve 3 (blue) of Flow-Stirrer cross-correlation (here, the small oscillation is related to electromechanical asymmetries of the stirrer). Finally, the Stirrer-TIP cross-correlation ---curve 2 (black)--- is antisymmetric, showing that the same amount of energy that the stirrer takes from the power supply at negative time lags is later released to the flow, at positive time lags.

In water, the energy transfer dynamics (for one stirrer, as before) is clearly different, as can be seen in the lower plots in Figure \ref{Fig_6}. In (c), the TIP and Flow power PDFs are nearly coincident and Gaussian, whereas the Stirrer power PDF is quite narrow and nearly symmetric, with negative skewness. Thus, the stirrer has a minor role in the whole dynamics. This is confirmed by the curves on subplot (d): the Stirrer-TIP cross-correlation, curve 2 (black) is nearly the opposite of the Flow-stirrer cross-correlation (curve 3 [blue]), the latter completely lacking the big dip present on the corresponding curve on subplot (b): in this case there is practically no energy storage in the stirrer. Moreover, in this case the Flow-TIP cross-correlation is  nearly symmetric, and the time lag of the peak is
\begin{equation}
T_\mathrm{w}=11.3~\mathrm{ms}.\label{TlagW}
\end{equation}
Thus, although the stirrers rotate approximately four times slower than in air, at similar Reynolds number the energy transfer dynamics in water is about three times faster. In terms of dimensionless time, we have
\begin{equation}
\frac{T_\mathrm{a}'}{T_\mathrm{w}'}=\frac{\overline{\Omega}_\mathrm{a} T_\mathrm{a}}{\overline{\Omega}_\mathrm{w} T_\mathrm{w}}\approx 12,
\label{trtime}
\end{equation}
so that at similar Reynolds numbers the energy transfer dynamics is about twelve times faster in water than in air, in dimensionless time. Of special interest is the comparison of the ratios of cutoff frequencies, given by equations (\ref{fcdm}) and (\ref{fcdml}). The reciprocal of this frequency is related to the beginning of the time scales in which the interaction between the stirrer and the neighboring turbulent flow undergoes a transition; from a state in which the stirrer simply follows the spatially averaged flow rotation, to the state where, for $f\gg f_\mathrm{c}$, it no longer responds to the fluctuating torque. From the spectra in water and air, and a linear fit of the frequency response obtained from equation (\ref{MEqW}) for each case, we have
\begin{equation}
\frac{f_\mathrm{c}^\mathrm{w}}{f_\mathrm{c}^\mathrm{a}}=\frac{0.627}{0.282}\approx 2.2,\label{fcdme}
\end{equation}
while the same ratio for the dimensionless frequencies is
\begin{equation}
\frac{f_\mathrm{c}^\mathrm{w}/\overline{\Omega}_\mathrm{w}}{f_\mathrm{c}^\mathrm{a}/\overline{\Omega}_\mathrm{a}}\approx 8.2.\label{fcdmle}
\end{equation}
These ratios reveal how much the stirrer response to the faster fluctuations of the torque exerted by the flow is enhanced when the fluid is water. As can be seen, these experimental values are almost six times smaller than those in equations (\ref{fcdm}) and (\ref{fcdml}). The results given in equations (\ref{fcdme}) and (\ref{fcdmle}), as compared with the estimates in equations (\ref{fcdm}) and (\ref{fcdml}), evidence a dramatic failure of the assumptions behind their obtention.

Along with the findings on the probability density functions and the energy transfer dynamics, the latter results tell us that there are deep differences in both, the flow and the energy transfer dynamics, when water replaces air in turbulent von K\'arm\'an flows. It is worth to stress that these differences cannot be ascribed to small differences in geometry or Reynolds number, because many preceding experiments with noticeable differences in geometry and Reynolds numbers have shown that the statistical properties of the injected power, in water and air separately, are fairly invariant. Here we see that there are clear discrepancies between the results obtained by assuming the validity of the weak similarity principle, when the angular speed $\Omega$ has small fluctuations, and those given by the experiments. Of course, we cannot claim that our experiments in water and air have exactly the same Re, or that the geometries are strictly similar. Instead, we can state that the differences that can be found in the experimental setups would very unlikely explain the discrepancies found in this study, when water replaces air in a von K\'arm\'an swirling flow.

\section{Conclusions}
\label{Conc}

Summarizing, we have found that running in constant torque mode, von K\'arm\'an swirling flows differ when the working fluid is water instead air. We ascribe this to the huge difference in densities between these two fluids, which in the case of water leads to a stronger coupling between the flow and the stirrer. Accordingly, the stirrer follows more closely the fluctuations of the torque exerted by the flow, which in turn leads to a nearly missing region with scaling $f^{-2}$ in the spectrum. Therefore, the dynamics issued from the interaction between the flow and the stirrers in each case produces dissimilar flows, which prevents thinking about this problem in terms of the hydrodynamic similarity principle, valid when geometrically similar setups run at equal Reynolds Numbers. This view is reinforced when we look at the ratio between the rms amplitude of the angular speed fluctuations and its mean value. In water, the ratio $\widetilde{\Omega}_w^{rms}/\overline{\Omega}_w \approx 4\times 10^{-2}$ is only five times greater than $\widetilde{\Omega}_a^{rms}/\overline{\Omega}_a \approx 8\times 10^{-3}$ in air, despite the fact that the water/air density ratio is $\rho_w/\rho_a\approx 830$, that is, about $200$ times larger. The other scales that we need to consider are: the scale factor between the experimental devices for air and water, which is $\lambda \approx 2$; and the moment of inertia of the stirrers used in air and water, which are $J_a\approx 4.7 \times 10^{-3}$~kg~m$^2$ and $J_w\approx 2 \times 10^{-3}$~kg~m$^2$ respectively, giving a ratio $J_a/J_w = 2.45$. If the similarity principle was valid for these two flows, then a linear scaling for similar flows gives a value $(\rho_w/\rho_a)(J_a/J_w)/\lambda \approx 10^3$ for the ratio $(\widetilde{\Omega}_w^{rms}/\overline{\Omega}_w)/(\widetilde{\Omega}_a^{rms}/\overline{\Omega}_a)$, which is $200$ times larger than the value obtained in the experiment. This huge discrepancy clearly indicates that if water is used in a given experimental setup, at equal Reynolds number the flow will not be similar to the flow obtained when the fluid is replaced by air, when the stirrers are driven at constant torque. This is not shocking at all: the difference is due to the difference in the response of the stirrers to the flow stresses. Note that these discrepancies are related only to fluctuations. For mean values, the similarity principle seems to work fairly well. For example, the ratio in equation (\ref{RWA}), which can be expressed in terms of mean values, differs from the experimental value only by $0.54\%$. Whether this is always the case when one is dealing with fully turbulent flows, remains an open question. From the outcomes reported here, it appears that high precision measurements would be necessary to obtain a response.\\

The consequences that these findings could have for studies using scale models will depend on how relevant turbulent fluctuations are to the system of interest. For example, if the air drag on a big truck traveling at $100$~km/h is to be studied using a small scale model in a water tunnel, possibly huge discrepancies in the relative rms amplitude of some magnitudes could be found if the similarity principle is naively applied. The reason is that these fluctuations will depend on the response of the scale model to the turbulent component of the flow, as our results suggest. Nonetheless, mean values will scale accordingly to the similarity principle fairly well, which may appear contradictory. This happens because when the turbulent component of the flow matters, the full 3D dynamic response of the body ---the truck in our example--- must be considered. In other words, in addition to scaling geometric and hydrodynamic parameters by using the similarity principle, an appropriate equivalent of equations (\ref{DynEq:3}) or (\ref{LangEq:1}) must be considered, in order to properly scale the mechanical parameters of the model; namely mass, supports' compliances, damping factors, and main moments of inertia.\\

We have to stress that none of the PDFs of total injected power obtained in the experiments reported here is Gaussian. The closest ones to a Gaussian are those for fluctuation of the angular speed for individual stirrers in water, which nevertheless have positive skewness, and the PDF of the reactive power for a stirrer in air, which has also positive skewness. These results are consistent with the findings by Titon and Cadot in water at constant torque: Figures 6(b) and 7(b) of their article\cite{TitCad03} display PDFs that clearly have positive skewness. A question remains: in which case the weak version of the similarity principle could be valid in von K\'arm\'an flows? The qualitative coincidence between the torque spectrum at constant angular speed\cite{LabPinFau96} and the deconvolved version\cite{LabBus12} ---at frequencies lower than $f_\mathrm{c}$--- obtained from fluctuations of angular speed in the constant external torque mode gives a clue. If we are able to run geometrically similar setups at equal Reynolds numbers using constant angular speed, then the PDFs of torque in air and water should be similar. The problem is that constant speed in water means using servo-controllers capable of keeping the angular speed of the stirrers well below the fluctuations of about $4$\% measured at constant torque. In other words, perhaps constant speeds within $0.04$\% or better would be necessary, along with stirrers having extremely low inertia.

Finally, in our experiment with air the Mach number is $M\lesssim0.05$, so that we do not expect that air compressibility plays a role in this phenomenon. Nevertheless, an experiment is being planned to test this possibility.

\begin{acknowledgments}
The authors gratefully acknowledge Lautaro Vergara and Ulrich Raff for their careful reading of the manuscript and valuable comments. Financial support for this work was provided in part by FONDECYT under project No. $1090686$ and DICYT-USACH under project No. $041231$LM. A. S. gratefully acknowledges financial support from CONICYT's \textit{Programa de Formaci\'on de Capital Humano Avanzado} and the fellowship from \textit{Direcci\'on General de Graduados} of Universidad de Santiago de Chile.
\end{acknowledgments}

\appendix

\section{Removal of cogging noise}
\label{Ap_A}

Two types of electric motors were used for the experiments reported here. Pancake servo motors were used with air, and universal motors with water. Pancake motors have no iron in their armatures, so that they have a very low inductance, run smoothly in their rated angular speed range, and can deliver their rated torque almost independently of the angular speed. They use permanent magnets to produce the stator magnetic field. Universal motors have winding in the stator to produce the stator field, and a cylindrical core in their armature. The latter is made of laminated iron, with a number of slots to allocate the windings. The slots can be helical or, most often, parallel to the axle. In the latter case, it happens that when the armature is in an angular position that minimizes the reluctance of the magnetic circuit, formed by the stator with its poles and the armature core, there appears a retentive torque which tends to anchor the armature in such angular position. This is because at these angles the magnetic flux density reaches a maximum. This condition is reproduced each time the angle of the armature advances by one slot. Thus, with $N_s$ slots there will be $N_s$ positions per turn where the armature will tend to become anchored. Note that the armature core can be seen as a cylinder with protuberances and slots in-between: hence the term ``cogging'' for this effect. On the other hand, the armature of a universal motor is highly inductive, so that its torque is degraded at high angular speeds. The interested reader can find a good introduction to electric motors in Ref. \onlinecite{Hugues06}.

A question that can be raised is: can inexpensive universal motors be used for this type of experiments? The answer is: it depends on the operating conditions.
\begin{figure}[t]
\centering \vspace{-0.5cm} \hspace{-0.2 cm}
\includegraphics[width=1.00\textwidth]{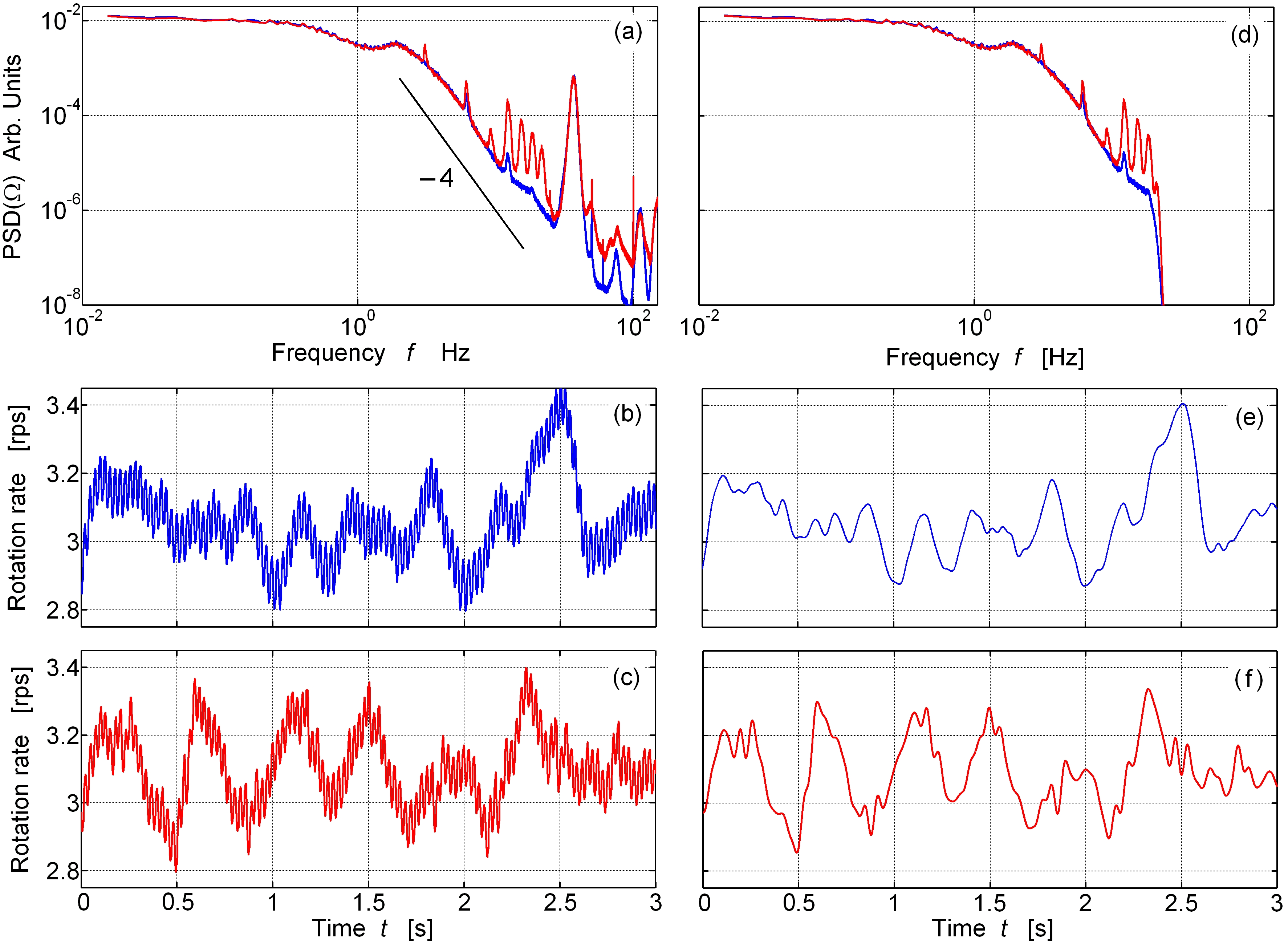}
\caption{(Color online) Rotation rate fluctuations of the stirrers in the device shown in Figure~\ref{Fig_1}. In the subplot (a), spectra showing roll-off regions with slope $s=-4$ can be seen. Subplots (b) and (c) display short records of the corresponding rotation rate signal. On the right half of subplot (a), a number of peaks appear, related to motor asymmetries. The largest peak is due to motor cogging (see text). Subplot (d) displays the spectra of the same signals, after processing them with a low-pass filter with cutoff frequency $f_c=19$~Hz. Subplots (e) and (f) show the effect of the filter on the signals displayed on the left column. The signal components related to the remaining peaks at frequencies below $f_c$ in the red spectrum have an amplitude too small to be seen on the subplot (f). Note that, after filtering, still about one decade of the region with slope $s=-4$ is preserved. Horizontal and vertical scales in plots of similar type are the same.} \label{Fig_7}
\end{figure}

In our experiment in water, the rotation rate is $\approx 3$~rps, which is well below the normal operating speed $> 200$~rps, typical of these motors. Thus, the reduction of torque at high speeds is not a concern. In addition, in this experiment the motors work with constant armature current, which at low angular speeds compensates for the armature inductance effects. The stator windings are powered by an independent, constant current (the field current), so that the mean stator magnetic field is constant. With respect to cogging, the number of slots in this case is $N_s=12$, which gives cogging frequencies in a narrow band centered at $f_\mathrm{cog}\approx 37$~Hz. This is near the end of the frequency band where the turbulence related fluctuations of the stirrers' angular speed take place. Thus, by using a low-pass filter the cogging noise can be easily removed. An additional bonus reduction of this noise comes from the necessity of large currents to obtain the high level of torque required by this experiment. The flux density must be increased by utilizing a field current well beyond the rated value for these motors. As a consequence, the working point of the magnetic flux is located deep in the saturation region, which reduces the changes in the flux density due to the armature motion. Normally, operation under these conditions should result in burned windings. To prevent such outcome, we implemented a powerful forced air cooling, which allowed continuous operation without overheating. Thus, under the conditions previously described, universal motors can give satisfactory results in this type of experiments when a better option in not available.

The cogging effect can be seen in Figure \ref{Fig_7}. Figure \ref{Fig_7}~(a) displays the rotation rate spectra of both disks: left, with few peaks in the high frequency zone (blue); and right, having a greater number of peaks (red). Below, the plots (b) and (c) display samples of the corresponding signals in the time domain. The rapid oscillations superimposed to the slow variations in both plots clearly illustrate the cogging effect. In the Figure \ref{Fig_7}~(d) the same spectra are displayed after applying a low pass filter. Notice that in the (red) spectrum three peaks before the cutoff frequency still remain. This noise seems to be related to asymmetries of the armature of this motor, and its amplitude is lower than that of the main peak. In the time domain, plots (e) and (f), we see that the rapid oscillation due to the cogging effect is eliminated. The red signal still contains some noise related to the peaks in the plot (d), but it is not detectable at the resolution level of plot (f). With the previous treatment, the signals obtained in the experiment with water are appropriate for the calculation of probability density functions.

To calculate correlation functions, a different signal treatment was used. To isolate the fundamental component of the cogging noise, a narrow band-pass filter tuned to the main peak frequency was used. Then this signal was subtracted from the main signal, adjusting its amplitude in order to minimize the peak height. Given that this signal component is modulated in both, amplitude and frequency, it was also necessary to adjust the filter bandwidth. Thus, the peak in the spectrum was also minimized with respect to the latter parameter. The cleaned signal allowed the calculation of inverse Fourier transforms on a wider interval of frequencies. Given that the spectral components of the whole signal fall a little more than six decades at $f\approx 100$~Hz, see Figure \ref{Fig_7}(a), the subtraction procedure previously described allowed a $[-100,100]$~Hz frequency window for the calculation of cross-correlation functions.

\section{Constant torque vs constant angular speed}

\subsection{Constant torque}

As noted in Section \ref{Intro}, the angular speed fluctuations of the stirrers in a von K\'arm\'an swirling flow are governed entirely by their interactions with the flow. The equation governing the motion of a stirrer is,
\begin{equation}
J \frac{d \Omega}{d t} +  \gamma_\mathrm{_M}\Omega + \rho R^5\overline{\tau}'\Omega^2 = \tau_\mathrm{_M} + \widetilde{\tau}.
\label{BE:1}
\end{equation}
Here we have omitted the weak torque related to the interaction of the flow with the inner surface of the container. This torque softly forces the mean global angular speed of the fluid to zero, and is not relevant in the analysis of the angular speed of the stirrers. To simplify our analysis, we will assume that all the variables are dimensionless, neglect the motor losses, and set the magnitude of every parameter to one. By writing $\Omega=\overline{\Omega}+\widetilde{\Omega}$,  we can obtain the equation of motion for the angular speed fluctuations. Keeping only linear terms in $\widetilde{\Omega}$, we get
\begin{equation}
\frac{d \widetilde{\Omega}}{d t} + \overline{\Omega}^2 + 2\overline{\Omega}\widetilde{\Omega} = \tau_\mathrm{_M} + \widetilde{\tau}.
\label{BE:2}
\end{equation}
Note that in equation (B2) the constant torque delivered by the motor, $\tau_\mathrm{_M}$, is balanced by the mean torque, $\overline{\Omega}^2$, exerted by the flow. Therefore, the motion equation for $\widetilde{\Omega}$ reduces to
\begin{equation}
\frac{d \widetilde{\Omega}}{d t} + 2\overline{\Omega}\widetilde{\Omega} = \widetilde{\tau}.
\label{BE:3}
\end{equation}
This is a first order Langevin equation for $\widetilde{\Omega}$, where the forcing term is the fluctuating part of the torque exerted by the flow: $\widetilde{\tau}$. The quantity $\widetilde{\tau}$ contains all the information about the turbulent flow that we can gather from its interaction with the stirrer, through the measurement of $\widetilde{\Omega}$. Of course, a definition of $\overline{\Omega}$ must be feasible for this simple equation to be valid. Now, given $\overline{\Omega}$, we can obtain $\widetilde{\tau}$ as the deconvolution of $\widetilde{\Omega}$. In the frequency domain, equation (B3) reads
\begin{equation}
2\pi i f \widetilde{\Omega}(f) +  2\overline{\Omega}\widetilde{\Omega}(f) = \widetilde{\tau}(f),
\label{BE:4}
\end{equation}
where $f$ is the frequency in Hz, and the functions with argument $f$ lie in the Fourier domain. The frequency response is, in this case
\begin{equation}
H(f)=\frac{\widetilde{\Omega}(f)}{\widetilde{\tau}(f)} = \frac{(2\pi f_\mathrm{c})^{-1}}{1+if/f_\mathrm{c}},
\label{BE:5}
\end{equation}
where
\begin{equation}
f_\mathrm{c}=\frac{\overline{\Omega}}{\pi},
\label{BE:6}
\end{equation}
is a cutoff frequency, which scales linearly with the mean angular speed $\overline{\Omega}$. In dimensional variables, and taking into account the motor losses, the cutoff frequency is
\begin{equation}
f_\mathrm{c}=\frac{\gamma_{_\mathrm{M}} + 2\rho R^5\overline{\tau}'\overline{\Omega}}{2\pi J},
\label{BE:7}
\end{equation}
Given the function $H(f)$, we can obtain $\widetilde{\tau}(t)$ from the measured $\widetilde{\Omega}(t)$ simply through
\begin{equation}
\widetilde{\tau}(t) = \mathfrak{F}^{-1} \biggl[\frac{\widetilde{\Omega}(f)N(f)}{H(f)}\biggr],
\label{BE:8}
\end{equation}
where $\mathfrak{F}^{-1} [~~]$ is the inverse Fourier transform and $N(f)$ is a filter designed to manage the divergence of $1/H(f)$ when $f\rightarrow\infty$, and suppress the noise at frequencies above the useful frequency band.

As can be seen, the experiment in which the electric motors are driven at constant torque allows a very simple analysis of the results. As the angular speed dynamics is that of a first order system, artifacts affecting the data acquisition/processing and/or the system dynamics could hardly be found. On the other hand, one has to be aware that in this driving mode, there is no independence between the flow dynamics and the stirrers motion. As noted in Section \ref{Intro}, a complex nonlinear dynamics is hidden in the forcing function $\widetilde{\tau}(t)$. Nevertheless, from this function, we can obtain valuable information about the structure of the turbulent flow.

\subsection{Constant angular speed}

Now let us consider a von K\'arm\'an swirling flow setup wanted to operate at constant angular speed. As before, we will neglect the motor losses, set the magnitude of all the parameters equal to one, and use dimensionless variables. Equation (\ref{BE:1}), which governs a stirrer dynamics, becomes
\begin{equation}
\frac{d \Omega}{d t} +  \Omega^2 = \tau_\mathrm{_M} + \widetilde{\tau},
\label{BE:9}
\end{equation}
where the only change is that the torque provided by the motor, $\tau_\mathrm{_M}$, is no longer constant: it must be determined by a servo controller in order to keep the angular speed constant, that is, make $d\Omega / dt=0$. In other words, ideally the controller must provide a torque $\tau_\mathrm{_M} = \overline{\tau}_\mathrm{_M} + \widetilde{\tau}_\mathrm{_M}$ such that
\begin{equation}
\overline{\tau}_\mathrm{_M} = \overline{\Omega}^2~~~\mathrm{and}~~~\widetilde{\tau}_\mathrm{_M} = -\widetilde{\tau}.
\label{BE:10}
\end{equation}

When the motor is powered by a voltage-controlled voltage source, it is necessary to take into account the inductance of the armature windings, which increases by one the order of the system. So, it is better to use a voltage-controlled current source, to keep the order of the system as low as possible. In this case the current source, which delivers the armature current $I_\mathrm{_A}$, directly determines $\tau_\mathrm{_M}=K I_\mathrm{_A}$, where $K$ is the motor's torque constant. Thus, assuming no limit to the current source compliance, we can make the controller output equal to the torque delivered to the stirrer.

Usually, a proportional-integral-derivative (PID) controller is used to set the speed at some reference value $\Omega_\mathrm{r}$. Assuming that the angular speed $\Omega$ can be measured instantaneously by some device, we can write the angular speed error as $e=\Omega_\mathrm{r}-\Omega$. For the moment, we will allow $\Omega_\mathrm{r}$ to change. Then, the output of the PID controller is
\begin{equation}
\tau_\mathrm{_M} = P e + I\int{e}dt + D\frac{d e}{dt},
\label{BE:11}
\end{equation}
where the parameters $P$, $I$, and $D$ are the proportional, integral, and derivative gains, respectively. By inserting this torque in (\ref{BE:9}), and expanding $e$, we obtain the system's dynamic equation:
\begin{equation}
\frac{d \Omega}{d t} +  \Omega^2 = P(\Omega_\mathrm{r}-\Omega)+I\int(\Omega_\mathrm{r}-\Omega)dt+D\frac{d(\Omega_\mathrm{r}-\Omega)}{dt} + \widetilde{\tau}.
\label{BE:12}
\end{equation}
In the steady state, we can expand the angular speed as $\Omega=\overline{\Omega}+\widetilde{\Omega}$. After some algebra, and retaining terms up to first order in $\widetilde{\Omega}$, we obtain
\begin{equation}
\frac{d\widetilde{\Omega}}{d t} + \overline{\Omega}^2+2\overline{\Omega}\widetilde{\Omega} = P(\Omega_\mathrm{r}-\overline{\Omega})-P\widetilde{\Omega}+ I\int(\Omega_\mathrm{r}-\overline{\Omega})dt-I\int{\widetilde{\Omega}}dt+D\frac{d(\Omega_\mathrm{r}-\widetilde{\Omega})}{dt} +\widetilde{\tau}.
\label{BE:13}
\end{equation}
With constant $\Omega_\mathrm{r}$, $\overline{\Omega}\rightarrow\Omega_\mathrm{r}$, and the term $\overline{\Omega}^2$ is balanced by $I\int(\Omega_\mathrm{r}-\overline{\Omega})dt$, so that
\begin{equation}
\frac{d\widetilde{\Omega}}{d t} + 2\overline{\Omega}\widetilde{\Omega} = \underbrace{-P\widetilde{\Omega}+ -I\int{\widetilde{\Omega}}dt - D\frac{d\widetilde{\Omega}}{dt} }_{\displaystyle =\widetilde{\tau}_\mathrm{_M}}+\widetilde{\tau}.
\label{BE:14}
\end{equation}
This equation gives the time evolution of the fluctuating part of the angular speed, $\widetilde{\Omega}$, under the joint action of the torque provided by the PID controller, $\widetilde{\tau}_\mathrm{_M}$, and the fluctuating part of the flow torque, $\widetilde{\tau}$. From equation (\ref{BE:14}) we can obtain $\widetilde{\tau}$ as a function of the angular speed fluctuations, and the controller parameters:
\begin{equation}
\widetilde{\tau}=(1+D)\frac{d\widetilde{\Omega}}{d t} + (2\overline{\Omega}+P)\widetilde{\Omega} + I\int{\widetilde{\Omega}}dt,
\label{BE:15}
\end{equation}
which in the Laplace domain reads
\begin{equation}
\widetilde{\tau}(s)=\biggl[(1+D)s +  (2\overline{\Omega}+P) + \frac{I}{s}\biggr]\widetilde{\Omega}(s).
\label{BE:16}
\end{equation}
Therefore, when a PID controller is used, the relationship between fluctuations of angular speed and fluctuation of flow torque is
\begin{equation}
\widetilde{\Omega}(s)=\frac{s}{(1+D)s^2 + (2\overline{\Omega}+P)s + I}~\widetilde{\tau}(s).
\label{BE:17}
\end{equation}
In the Laplace domain, the term $\widetilde{\tau}_\mathrm{_M}$ in equation (\ref{BE:14}) translates  into
\begin{equation}
\widetilde{\tau}_\mathrm{_M}(s)=-\biggl(P + \frac{I}{s} + Ds\biggr)\widetilde{\Omega}(s),
\label{BE:18}
\end{equation}
so that, from equations (\ref{BE:17}) and (\ref{BE:18})
\begin{equation}
\widetilde{\tau}_\mathrm{_M}(s)=-\frac{Ds^2 +Ps + I}{(1+D)s^2 + (2\overline{\Omega}+P)s + I}\widetilde{\tau}(s).
\label{BE:19}
\end{equation}
From here, it can be shown that the error in the torque measurement, as a function of $\widetilde{\Omega}$, is
\begin{equation}
\varepsilon_{\tau}(s)=\widetilde{\tau}_\mathrm{_M}(s)+\widetilde{\tau}(s) = (s+2\overline{\Omega})\widetilde{\Omega}(s).
\label{BE:20}
\end{equation}
Equation (\ref{BE:17}) tells us that a PID controller cannot keep a perfectly constant angular speed, because $\widetilde{\Omega}(s)\neq 0$. In the frequency domain it reads
\begin{equation}
\widetilde{\Omega}(f)=\frac{2 \pi i f}{I - 4(1+D)\pi^2 f^2 + 2\pi i(2\overline{\Omega}+P)f}~\widetilde{\tau}(f),
\label{BE:21}
\end{equation}
so that in the low frequency band, $f\rightarrow 0$, the controller can keep $\widetilde{\Omega}(f)\approx 0$. But at finite frequencies the error increases, and could reach a maximum when $I \approx 4(1+D)\pi^2 f^2$ if non optimized values for $P$, $I$, and $D$ are used. When  $f\rightarrow \infty$ the error decreases again. Of course, a sharp resonance at $f\approx \sqrt{I/4\pi^2(1+D)}$ can be avoided with proper parameter adjustment, but even with optimal values, an error of possibly non negligible amplitude may persist in a more or less wide band of frequencies. Within this band, the stirrer angular speed will fluctuate under the action of the torque exerted by the flow, so that the torque measurement will be contaminated by the angular speed fluctuations. This can be derived from the equation (\ref{BE:19}): the motor torque $\widetilde{\tau}_\mathrm{_M}$ set by the controller cannot exactly mirror the torque $\widetilde{\tau}$ exerted by the flow. From equation (\ref{BE:20}) we have
\begin{equation}
\widetilde{\tau}_\mathrm{_M}(s) = -\widetilde{\tau}(s) + (s+2\overline{\Omega})\widetilde{\Omega}(s)~,
\label{BE:22}
\end{equation}
where the contamination of the torque measurement by angular speed fluctuations is explicitly displayed. Note that it gets worse at higher values of $\overline{\Omega}$. The accurate measurement of power fluctuation at constant angular speed requires a vanishing amplitude in the last term of equation (\ref{BE:22}), in which case the desired relation $\widetilde{\tau}_\mathrm{_M}=-\widetilde{\tau}(s)$ could be achieved. However, when a PID controller is used in this context, there will be always a mismatch between the zeros of the polynomials in the numerator and denominator of the equation (\ref{BE:19}), no matter the values of the parameters $P,I,D$. On the other hand, the rational function in this equation, which is the steady state transfer function $G(s)$ linking $\widetilde{\tau}(s)$ and $\widetilde{\tau}_\mathrm{_M}(s)$,
\begin{equation}
G(s)=\frac{\widetilde{\tau}_\mathrm{_M}(s)}{\widetilde{\tau}(s)}=-\frac{Ds^2 +Ps + I}{(1+D)s^2 + (2\overline{\Omega}+P)s + I}~,
\label{BE:23}
\end{equation}
has the mean angular speed $\overline{\Omega}$ as one of its parameters. Thus, a change in $\overline{\Omega}$ changes the overall system response. Although this idealized model shows that an approximate matching of the zeros in the numerator and denominator of $G (s) $ can be obtained if one chooses $D\gg1$ and $P\gg2\overline{\Omega}$, too big values in these parameters will inevitably generate instability problems arising from bandwidth limitations in one or more of the controller components. In addition, high frequency noise problems related to large values of $D$ would arise.

We must stress that in no way we are suggesting that a PID controller is useless for this application. With appropriate tuning, the error $\varepsilon_\tau(s)$ can in principle be reduced to an acceptable level. Moreover, finding the dependence of the parameters $P$, $I$, and $D$ on $\overline{\Omega}$, using as control criteria the minimization $\varepsilon_\tau(s)$ for each value of $\overline{\Omega}$, should allow the design of  an optimal auto-tuning PID controller.\\

Summarizing, both methods have pros and cons when it comes to the measurement of power fluctuations. On the one hand, the constant torque mode may be the simplest option, although the measurement of torque (via deconvolution) or angular speed has always a hidden component related to the interplay between the angular acceleration and the changes that it produces in the turbulent flow. Yet, in the experiments reported here this mode revealed how the energy transfer dynamics changes due to a large change in the fluid density. On the other hand, the constant speed mode requires a careful tuning of the servo-controller, in order to avoid the contamination of torque measurement by fluctuations of the angular speed, along with the effects that the angular acceleration have on the turbulent flow structure. In principle, a well tuned PID controller should minimize such effects, but a rigorous assessment of the overall system error is crucial. Of course, there are alternative control strategies that, specifically for studies in von K\'arm\'an swirling flows, could have better performance than the well known PID controller.

The reader not familiar with control theory will find an excellent introduction to feedback principles and control systems in Ref. \onlinecite{Bech05}.


%

\end{document}